\documentclass[twocolumn]{aastex631}

\usepackage{multirow}
\usepackage{mathtools}
\usepackage[]{hyperref}
\usepackage{subfigure}

\hypersetup{
    colorlinks=true,
    linkcolor=blue,
    filecolor=magenta,      
    urlcolor=cyan,
    pdftitle={Overleaf Example},
    pdfpagemode=FullScreen,
    }

\begin{document}

\title{The NANOGrav 12.5-Year Data Set: Dispersion Measure Mis-Estimation with Varying Bandwidths}

\author[0000-0002-5176-2924]{Sophia V. Sosa Fiscella}
\affiliation{School of Physics and Astronomy, Rochester Institute of Technology, Rochester, NY 14623, USA}
\affiliation{Laboratory for Multiwavelength Astrophysics, Rochester Institute of Technology, Rochester, NY 14623, USA}
\author[0000-0003-0721-651X]{Michael T. Lam}
\affiliation{SETI Institute, 339 N Bernardo Ave Suite 200, Mountain View, CA 94043, USA}
\affiliation{School of Physics and Astronomy, Rochester Institute of Technology, Rochester, NY 14623, USA}
\affiliation{Laboratory for Multiwavelength Astrophysics, Rochester Institute of Technology, Rochester, NY 14623, USA}
\author{Zaven Arzoumanian}
\affiliation{X-Ray Astrophysics Laboratory, NASA Goddard Space Flight Center, Code 662, Greenbelt, MD 20771, USA}
\author[0000-0003-4046-884X]{Harsha Blumer}
\affiliation{Department of Physics and Astronomy, West Virginia University, P.O. Box 6315, Morgantown, WV 26506, USA}
\affiliation{Center for Gravitational Waves and Cosmology, West Virginia University, Chestnut Ridge Research Building, Morgantown, WV 26505, USA}
\author[0000-0003-3053-6538]{Paul R. Brook}
\affiliation{Institute for Gravitational Wave Astronomy and School of Physics and Astronomy, University of Birmingham, Edgbaston, Birmingham B15 2TT, UK}
\author[0000-0002-6039-692X]{H. Thankful Cromartie}
\altaffiliation{NASA Hubble Fellowship: Einstein Postdoctoral Fellow}
\affiliation{Cornell Center for Astrophysics and Planetary Science and Department of Astronomy, Cornell University, Ithaca, NY 14853, USA}
\author[0000-0002-2185-1790]{Megan E. DeCesar}
\affiliation{George Mason University, resident at the Naval Research Laboratory, Washington, DC 20375, USA}
\author[0000-0002-6664-965X]{Paul B. Demorest}
\affiliation{National Radio Astronomy Observatory, 1003 Lopezville Rd., Socorro, NM 87801, USA}
\author[0000-0001-8885-6388]{Timothy Dolch}
\affiliation{Department of Physics, Hillsdale College, 33 E. College Street, Hillsdale, MI 49242, USA}
\affiliation{Eureka Scientific, 2452 Delmer Street, Suite 100, Oakland, CA 94602-3017, USA}
\author{Justin A. Ellis}
\altaffiliation{Infinia ML, 202 Rigsbee Avenue, Durham NC, 27701}
\author[0000-0002-2223-1235]{Robert D. Ferdman}
\affiliation{School of Chemistry, University of East Anglia, Norwich, NR4 7TJ, United Kingdom}
\author[0000-0001-7828-7708]{Elizabeth C. Ferrara}
\affiliation{Department of Astronomy, University of Maryland, College Park, MD 20742}
\affiliation{Center for Research and Exploration in Space Science and Technology, NASA/GSFC, Greenbelt, MD 20771}
\affiliation{NASA Goddard Space Flight Center, Greenbelt, MD 20771, USA}
\author[0000-0001-8384-5049]{Emmanuel Fonseca}
\affiliation{Department of Physics and Astronomy, West Virginia University, P.O. Box 6315, Morgantown, WV 26506, USA}
\affiliation{Center for Gravitational Waves and Cosmology, West Virginia University, Chestnut Ridge Research Building, Morgantown, WV 26505, USA}
\author[0000-0001-6166-9646]{Nate Garver-Daniels}
\affiliation{Department of Physics and Astronomy, West Virginia University, P.O. Box 6315, Morgantown, WV 26506, USA}
\affiliation{Center for Gravitational Waves and Cosmology, West Virginia University, Chestnut Ridge Research Building, Morgantown, WV 26505, USA}
\author[0000-0001-8158-683X]{Peter A. Gentile}
\affiliation{Department of Physics and Astronomy, West Virginia University, P.O. Box 6315, Morgantown, WV 26506, USA}
\affiliation{Center for Gravitational Waves and Cosmology, West Virginia University, Chestnut Ridge Research Building, Morgantown, WV 26505, USA}
\author[0000-0003-1884-348X]{Deborah C. Good}
\affiliation{Department of Physics, University of Connecticut, 196 Auditorium Road, U-3046, Storrs, CT 06269-3046, USA}
\affiliation{Center for Computational Astrophysics, Flatiron Institute, 162 5th Avenue, New York, NY 10010, USA}
\author[0000-0001-6607-3710]{Megan L. Jones}
\affiliation{Center for Gravitation, Cosmology and Astrophysics, Department of Physics, University of Wisconsin-Milwaukee,\\ P.O. Box 413, Milwaukee, WI 53201, USA}
\author[0000-0003-1301-966X]{Duncan R. Lorimer}
\affiliation{Department of Physics and Astronomy, West Virginia University, P.O. Box 6315, Morgantown, WV 26506, USA}
\affiliation{Center for Gravitational Waves and Cosmology, West Virginia University, Chestnut Ridge Research Building, Morgantown, WV 26505, USA}
\author[0000-0001-5373-5914]{Jing Luo}
\altaffiliation{Deceased}
\affiliation{Department of Astronomy \& Astrophysics, University of Toronto, 50 Saint George Street, Toronto, ON M5S 3H4, Canada}
\author[0000-0001-5229-7430]{Ryan S. Lynch}
\affiliation{Green Bank Observatory, P.O. Box 2, Green Bank, WV 24944, USA}
\author[0000-0001-7697-7422]{Maura A. McLaughlin}
\affiliation{Department of Physics and Astronomy, West Virginia University, P.O. Box 6315, Morgantown, WV 26506, USA}
\affiliation{Center for Gravitational Waves and Cosmology, West Virginia University, Chestnut Ridge Research Building, Morgantown, WV 26505, USA}
\author[0000-0002-3616-5160]{Cherry Ng}
\affiliation{Dunlap Institute for Astronomy and Astrophysics, University of Toronto, 50 St. George St., Toronto, ON M5S 3H4, Canada}
\author[0000-0002-6709-2566]{David J. Nice}
\affiliation{Department of Physics, Lafayette College, Easton, PA 18042, USA}
\author[0000-0001-5465-2889]{Timothy T. Pennucci}
\affiliation{Institute of Physics and Astronomy, E\"{o}tv\"{o}s Lor\'{a}nd University, P\'{a}zm\'{a}ny P. s. 1/A, 1117 Budapest, Hungary}
\author[0000-0002-8826-1285]{Nihan S. Pol}
\affiliation{Department of Physics and Astronomy, Vanderbilt University, 2301 Vanderbilt Place, Nashville, TN 37235, USA}
\author[0000-0001-5799-9714]{Scott M. Ransom}
\affiliation{National Radio Astronomy Observatory, 520 Edgemont Road, Charlottesville, VA 22903, USA}
\author[0000-0002-6730-3298]{Ren\'{e}e Spiewak}
\affiliation{Jodrell Bank Centre for Astrophysics, University of Manchester, Manchester, M13 9PL, United Kingdom}
\author[0000-0001-9784-8670]{Ingrid H. Stairs}
\affiliation{Department of Physics and Astronomy, University of British Columbia, 6224 Agricultural Road, Vancouver, BC V6T 1Z1, Canada}
\author[0000-0002-7261-594X]{Kevin Stovall}
\affiliation{National Radio Astronomy Observatory, 1003 Lopezville Rd., Socorro, NM 87801, USA}
\author[0000-0002-1075-3837]{Joseph K. Swiggum}
\altaffiliation{NANOGrav Physics Frontiers Center Postdoctoral Fellow}
\affiliation{Department of Physics, Lafayette College, Easton, PA 18042, USA}
\author[0000-0003-4700-9072]{Sarah J. Vigeland}
\affiliation{Center for Gravitation, Cosmology and Astrophysics, Department of Physics, University of Wisconsin-Milwaukee,\\ P.O. Box 413, Milwaukee, WI 53201, USA}




\begin{abstract}
Noise characterization for pulsar-timing applications accounts for interstellar dispersion by assuming a known frequency-dependence of the delay it introduces in the times of arrival (TOAs). However, calculations of this delay suffer from mis-estimations due to other chromatic effects in the observations. The precision in modeling dispersion is dependent on the observed bandwidth. In this work, we calculate the offsets in infinite-frequency TOAs due to mis-estimations in the modeling of dispersion when using varying bandwidths at the Green Bank Telescope. We use a set of broadband observations of PSR~J1643$-$1224, a pulsar with an excess of chromatic noise in its timing residuals. We artificially restricted these observations to a narrowband frequency range, then used both data sets to calculate residuals with a timing model that does not include short-scale dispersion variations. By fitting the resulting residuals to a dispersion model, and comparing the ensuing fitted parameters, we quantify the dispersion mis-estimations. Moreover, by calculating the autocovariance function of the parameters we obtained a characteristic timescale over which the dispersion mis-estimations are correlated. For PSR~J1643$-$1224, which has one of the highest dispersion measures (DM) in the NANOGrav pulsar timing array, we find that the infinite-frequency TOAs suffer from a systematic offset of $\sim22~\mu$s due to DM mis-estimations, with correlations over $\sim1$ month. For lower-DM pulsars, the offset is $\sim7~\mu$s. This error quantification can be used to provide more robust noise modeling in NANOGrav's data, thereby increasing sensitivity and improving parameter estimation in gravitational wave searches.
\end{abstract}



\section{Introduction} \label{sec:intro}
In 2010, the National Radio Astronomy Observatory (NRAO) launched the Green Bank Ultimate Pulsar Processing Instrument \citep[GUPPI;][]{2008SPIE.7019E..1DD}, a digital signal processor designed for pulsar observations with the 100-m Robert C. Byrd Green Bank Telescope (GBT). Its large bandwidth coherent de-dispersion observation modes constituted a significant improvement over previously available backends (see Table \ref{tab:observing_frequencies}), such as the Green Bank Astronomical Signal Processor \citep[GASP;][]{demorest}.


The advent of receivers with larger bandwidths allows for better estimates of the dispersion measure (DM). These estimates are obtained by precisely measuring the arrival times of pulsar emission as a function of radio frequency \citep[e.g.,][]{manchester1977pulsars,2002ASPC..278..251S,2004hpa..book.....L}. Therefore, sampling more frequencies \citep[up to a point, see e.g.,][]{2018ApJ...861...12L} allows for better modeling of interstellar dispersion and mitigates mis-estimations introduced by other frequency-dependent delays. Conversely, DM mis-estimations will be more pronounced in observations with narrowband receivers, due to sampling a smaller frequency space to model the dispersion of the signal. By comparing simultaneous observations in both frequency regimes, we can quantify the errors introduced by using a narrower frequency band.

Mis-estimations in the DM can arise from using narrowband frequency sampling \citep{2017MNRAS.464.2075S}, using incorrect temporal correlations between different channels due to uncertaintites in the ISM diffractive index \citep{2016PhDT.......200L}, the combination of asynchronously observed multifrequency data \citep{2015ApJ...801..130L}, or frequency-dependent DMs due to interstellar scattering \citep{2016ApJ...817...16C}. Such mis-estimations will cause red noise in the timing residuals \citep{2013MNRAS.429.2161K}. If unaccounted for, this noise can severely hinder the resulting timing precision \citep{2018ApJ...861...12L}. This is of special relevance for all high-precision pulsar timing experiments involving frequency-dependent effects, such as gravitational wave (GW) searches \citep{, NG15GWB,EPTAGWB,PPTAGWB,CPTAGWB}, calculating pulse broadening functions employing CLEAN deconvolution algorithms \citep{2023arXiv230606046Y}, monitoring interstellar scattering delays \citep{2021ApJ...917...10T}, and studying jitter in millisecond pulsars \citep{2019ApJ...872..193L}. Therefore, modeling and accounting for DM mis-estimations when using narrowband receivers is essential for providing realistic timing errors.

Observations with unaccounted errors due to DM mis-estimations are not suitable for high-precision timing experiments, so they are usually not included in such studies \citep[e.g.,][]{2018Natur.559...73A,EPTAGWB}. However, this approach reduces the available time baseline of pulsar observations, therefore decreasing our sensitivity to long-period gravitational waves. In this work, we propose an alternative approach: we perform narrow-bandwidth DM estimations using the GUPPI data set and compare the offset with the broader-bandwidth values to estimate and correct for the mis-estimations of the DM.

This type of analysis will be of special interest as we advance towards a new generation of wideband receivers. In particular, the introduction of the VErsatile GBT Astronomical Spectrometer \citep[VEGAS;][]{2012AAS...21944610B} can duplicate all the capabilities of the GUPPI backend, but also allows for wider instantaneous bandwidths of up to 3.8 GHz.

In Sec.~\ref{sec:context} we describe the different frequency-dependent delays affecting the signal propagation, and how they are incorporated into our timing model. In Sec.~\ref{sec:observations} we provide information on the data collection and reduction methods used for NANOGrav's observations with the GBT. In Sec.~\ref{sec:results}, we describe the main pulsar we analyzed in this work as a case study, PSR~J1643$-$1224. We also outline the methods in producing different timing residuals for isolating and quantifying the DM mis-estimations due to varying bandwidths and the time correlations in these variations. In Sec.~\ref{sec:conclusions} we summarize the results and implications of this work. Processed data products presented here are publicly available\footnote{\url{https://github.com/sophiasosafiscella/DM_misestimations}} as of the date this work is published.

\section{Methodology}

\subsection{Frequency-Dependent Delays} \label{sec:context}

As a radio signal is emitted at a pulsar and as it propagates through the ISM, it will encounter  various frequency-dependent delays. As a result, pulses with frequency $\nu$ will arrive at Earth's position at a time $t_{\nu}$ that is delayed with respect to the expected time $t_{\infty}$ for a signal of infinite frequency (i.e., assuming no chromatic effects). In this section we will describe the various frequency-dependent effects on the TOAs responsible for this delay, and how they are accounted for in our timing models.

For each pulsar, TOAs are calculated for all frequency channels recorded with a given receiver using a single standard template profile. However, pulse shapes vary with frequency \citep{Kramer_1998, Pennucci_2014} even with no intervening ISM,  so when compared against a single-frequency template this introduces small systematic  frequency-dependent perturbations in the TOAs. These changes are modeled as polynomials in log-frequency, described by the $\mathrm{FD}_{k}$ parameters as \citep{NANOGrav_9yr}

\begin{equation}\label{eq:FD_parameters}
    t_{\mathrm{PE}} = \sum_{k=1}^{n} \mathrm{FD}_{k} \log \left( \frac{\nu}{1~\mathrm{GHz}}\right)^{k}
\end{equation}

\noindent The number of terms needed varies for any given pulsar, but $n=2$ parameters suffice to describe the pulse evolution with frequency for PSR J1643$-$1223. There is no $k=0$ term because this would be a constant phase offset that is removed when the mean is subtracted to the timing residuals.

While propagating through the ionized plasma, the pulsar signal encounters ionized plasma and electron density variations along the way, resulting in a frequency-dependent index of refraction.
As a result, the radiation will suffer a first-order chromatic delay -- the interstellar dispersion -- which is the largest frequency-dependent effect due to the ISM. For a cold, unmagnetized plasma, a pulse observed at frequency $\nu$ is delayed compared to one at infinite frequency by an amount $t_{\mathrm{DM}} = K \times \mathrm{DM}/\nu^2$, where the dispersion measure ($\mathrm{DM}$) is the line-of-sight (LOS) integral of the free electron along the line of sight to a pulsar.  The DM can be quantified as $\mathrm{DM} = \int_{0}^{d} n_{e} (l) dl$, where $n_{e}$ is the free electron density along the line of sight $l$, and $d$ is the pulsar distance. The  dispersion constant is given by $K = e^2 / 2 \pi m_e c \simeq 4.149~\mathrm{ms}~\mathrm{GHz}^2 \mathrm{pc}^{-1}~\mathrm{cm}^3$.

Turbulent and bulk motions within the ISM, solar wind, differences in the relative velocity of the pulsar and the Earth, and stochastic variations from pulsar motion can cause the line of sight to sample electron-density fluctuations on a variety of scales  \citep{Lam2016, Cordes_1998, 1991ApJ...382L..27P}. The result is a DM that varies with time, changing on timescales of hours to years. To model these short-scale variations, we use a stepwise model for variation in DM, in which DM is allowed to
have independently varying values in time intervals. The time
intervals range in length from $0.5$ to $15$ days, depending on the
telescope and instrumentation. The offset from the globally fixed fiducial $\mathrm{DM}$ value is given by the epoch-dependent $\mathrm{DMX}$ parameters \citep[e.g.,][]{2017ApJ...841..125J, 2021ApJ...909..219S} The ensuing timing correction is given by $t_{\mathrm{DMX}_{i}} = K \times\mathrm{DMX}_{i}/\nu^2$, where $\mathrm{DMX}_{i}$ is the correction corresponding to the observing epoch $i$.

In addition to the $\nu^{-2}$ offsets due to dispersion\footnote{Angle-of-arrival variations from refraction will also yield $\nu^{-2}$ delays \citep{fc1990} but since they are entirely covariant with the dispersive correction, they are absorbed in the fit and we will ignore them further.}, there are also a variety of frequency-dependent effects for which the perturbations scale with radio frequency obeying a $\nu^{-\alpha}$ power law \citep[e.g.,][]{Lam2016}. This includes geometric perturbations such as delays due to incorrectly referencing the arrival time of the pulse at the solar system barycenter \citep[$\alpha=2$;][]{fc1990}. It also comprises pulse scattering, associated with the variable path length due to refraction which results in the signal reaching the observer along different geometrical paths \citep[e.g.,][]{Bansal_2019, 10.1093/mnras/stt989}. As a result, the pulse will arrive at the observer over a finite interval and it will be enveloped by a pulse broadening function. The thin-screen scattering model \citep[e.g.,][]{2017MNRAS.464.2075S} considers an isotropic homogeneous turbulent medium with a Gaussian \citep[$\alpha=4$;][]{1971ApJ...164..249L} or a Kolmogorov \citep[$\alpha=4.4$;][]{10.1093/mnras/220.1.19} distribution of inhomogeneities. Finally, interstellar scintillation will introduce a random component in the TOA delay whose variance is strongly frequency dependent. 
A simple model for the single-epoch TOA delay introduced by these chromatic effects is:

\begin{equation}\label{eq:chromatic}
    t_{\mathrm{C},\nu} = \sum_{i=1}^{N_c} C_{i} \nu^{-\alpha_i}
\end{equation}

\noindent where the $i = 1, N_c$ additional terms model chromatic TOA variations with
unique power-law spectral scalings $\alpha_i$. 

By incorporating all these effects into our timing model, we quantify the time of arrival (TOA) as a function of frequency $\nu$ as
\begin{equation}\label{eq:disp}
    t_{\nu, i} = t_{\infty, i} + \frac{K \times (\mathrm{DM} + \mathrm{DMX}_{i})}{\nu^2} + t_{\mathrm{PE}, \nu} + t_{\mathrm{C},\nu,i}
\end{equation}

\noindent where the subscript $i$ denotes the epoch.

In addition to these delays, there are non-power-law frequency-dependent timing effects that can modify the pulse arrival times. As previously stated, the DM is defined as the line-of-sight integral of the electron density. However, the line of sight can change as a function of frequency as a result of ray paths at different frequencies covering different volumes through the ISM \citep{2016ApJ...817...16C}. Therefore, DM itself is frequency-dependent, i.e., $t_{\rm DM} = K \times \mathrm{DM} (\nu)/\nu^2$, and this dependence will introduce timing errors that cannot be mitigated solely by increasing the observing bandwidth \citep{2018ApJ...861...12L}. Instead, in this analysis we fit only for a constant DM over the observation, and aim to quantify the mis-estimation in the DM that are introduced by other chromatic effects in the observation altering the DM fit. Other non-power-law frequency-dependent timing effects may be present at small amplitudes along specific lines of sight as well; for example, refraction through plasma-lens-like structures in the ISM is manifested in distinctive DM and geometric path variations that result in such delays \citep[see e.g., Eq. 17 in ][]{2017ApJ...842...35C}. These effects result in higher-order corrections in the measured TOAs that we can neglect for the precision required in this work.


\begin{figure}[ht!]
\epsscale{1.2}
\plotone{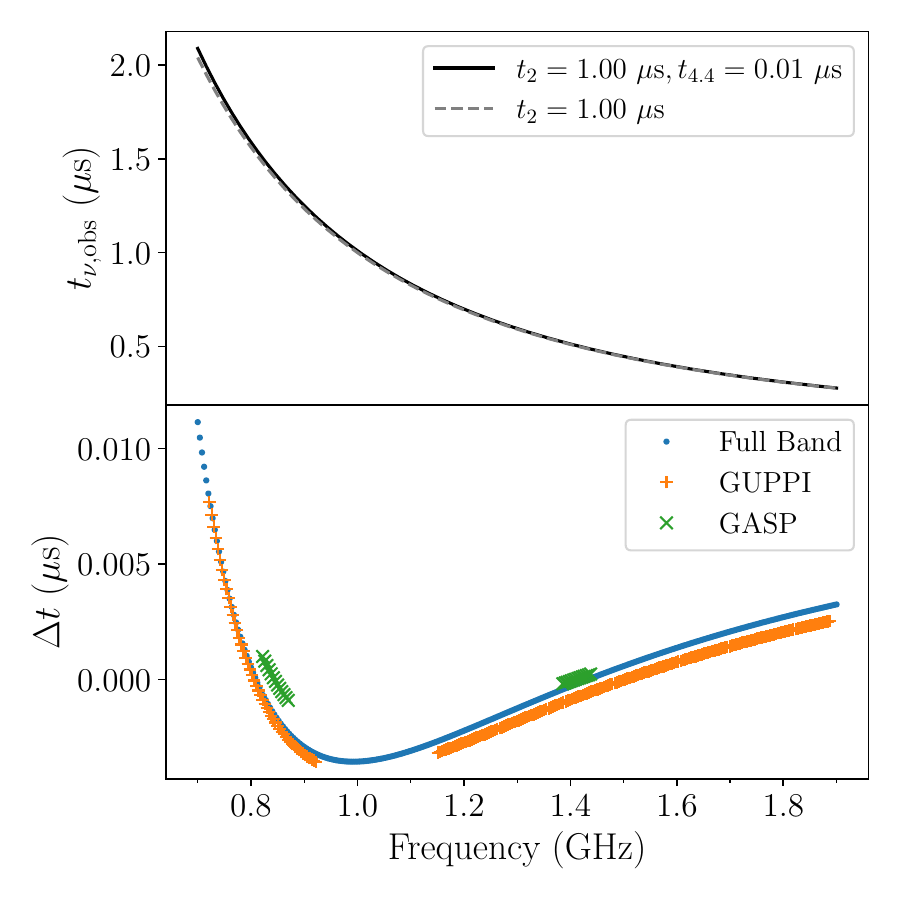}
\caption{Upper figure: an artificial set of TOAs, presenting a characteristic dispersion curve as a result of frequency-dependent delays. Lower figure: qualitative representation of the expected timing residuals when using GASP's bandwidth (green), GUPPI's bandwidth (orange), and the full range of frequencies (blue).
\label{fig:cartoon}}. 
\end{figure}

\subsection{Simulated Example}

\begin{deluxetable*}{lcccBccc}
\tabletypesize{\scriptsize}
\tablewidth{0pt} 
\tablecaption{\centering Observing Frequencies and Bandwidths. Source: \cite{2023ApJ...951L...9A}}\label{tab:observing_frequencies}
\tablehead{
\multirow{4}{2cm}{Telescope Receiver} & \multicolumn{3}{c}{GASP} & \multicolumn{3}{c}{GUPPI} \\
\cline{2-4}
\cline{6-8}
\colhead{} & \colhead{Data Span$^a$} & \colhead{\multirow{3}{2.0cm}{\centering Full Frequency Range$^b$ (MHz)}} & \colhead{\multirow{3}{1.5 cm}{\centering Usable Bandwidth$^c$ (MHz)}} & \colhead{} & \colhead{Data Span$^a$} & \colhead{\multirow{3}{2.0cm}{\centering \centering Full Frequency Range$^b$ (MHz)}} & \colhead{\multirow{3}{1.5cm}{\centering Usable Bandwidth$^c$ (MHz)}} \\
} 
\startdata 
Rcvr\_800 & $2004.6$--$2011.0$ & $792$-$884$ & $64$ & & $2010.2$--$2020.3$& $725$-$916$ & $180$ \\
Rcvr1\_2 & $2004.6$--$2010.8$ & $1340$-$1432$ & $64$ & & $2010.2$--$2020.3$& $1156$-$1882$ & $640$ \\
\enddata
\tablecomments{\\ $^a$Dates of instrument use. Observation dates of individual pulsars vary \\ $^b$ Typical values; some observations differed. Some frequencies were unusable owing to radio frequency interference.\\ $^c$ Approximate and representative values after excluding narrow sub-bands with radio frequency interference}
\end{deluxetable*}

In Fig.~\ref{fig:cartoon} we present a qualitative representation of the timing residuals (differences between the observed times of arrival and the predictions from the timing model) that would be obtained by applying a simple timing model to three sets of TOAs that cover either GASP's bandwidth, GUPPI's bandwidth, or the full range of frequencies from $0.7~\mathrm{GHz}$ to $1.9~\mathrm{GHz}$. For each backend, an artificial set of frequency-dispersed TOAs was fabricated by simplifying Eq.~\ref{eq:disp} as $t_{\mathrm{obs}}(\nu)=\mathrm{K \times DM} / \nu^{2} + t_{\mathrm{C}} /\nu^{4}$, using frequencies $\nu$ in the backend's bandwidth. We choose $\mathrm{K \times DM}=1~\mathrm{GHz}$ for the dispersion coefficient and $t_{\mathrm{C}}=0.01~\mathrm{GHz}$ for the chromatic coefficient. The ``observed" TOAs were used to fit a simple timing model $t_{\mathrm{pred}}(\nu) = a + b/\nu^2$ that only accounts for $\nu^{-2}$-delays in order to simulate the effects of having other chromatic effects being absorbed into the DM fit. This function is then evaluated at the same frequencies to obtain ``predicted" TOAs. By subtracting both sets of TOAs we obtain the residuals $\Delta t = t_{\mathrm{obs}} - t_{\mathrm{pred}}$ presented in the figure. We observe that the residuals vary significantly depending on the bandwidth of the backend that was used to take the observations. Effectively, a larger bandwidth allows for a larger sampling of the frequency space over which to model the dispersion. The GUPPI backend combines non-simultaneous observations around two frequency bands, one centered near 820 MHz and another near 1400 MHz, to cover most of the bandwidth from 0.7 GHz to 1.9 GHz. Even then, the resulting timing residuals differ from the estimation we would expect if the receiver covered the full band. As a result, we expect a bias in our measurements even when more advanced backends are used. 



\section{Pulsar backends and observations} \label{sec:observations}


In the present analysis we used observations from NANOGrav's 12.5-year data set release \citep{Nanograv_12yr} obtained using the GBT at the Green Bank Observatory in West Virginia, USA. Two radio receivers at separated frequency bands were used to perform the observations: one centered near $820$ MHz and another near $1400$ MHz (see Table~\ref{tab:observing_frequencies}). The observations were performed with a monthly cadence using both receivers. However, these two separate frequency ranges were not observed simultaneously. Instead, the observations were separated by a few days due to the need for a physical receiver change at that telescope. The typical observation duration was about 25 minutes.

\begin{figure*}[ht!]
\epsscale{1.0}
\plotone{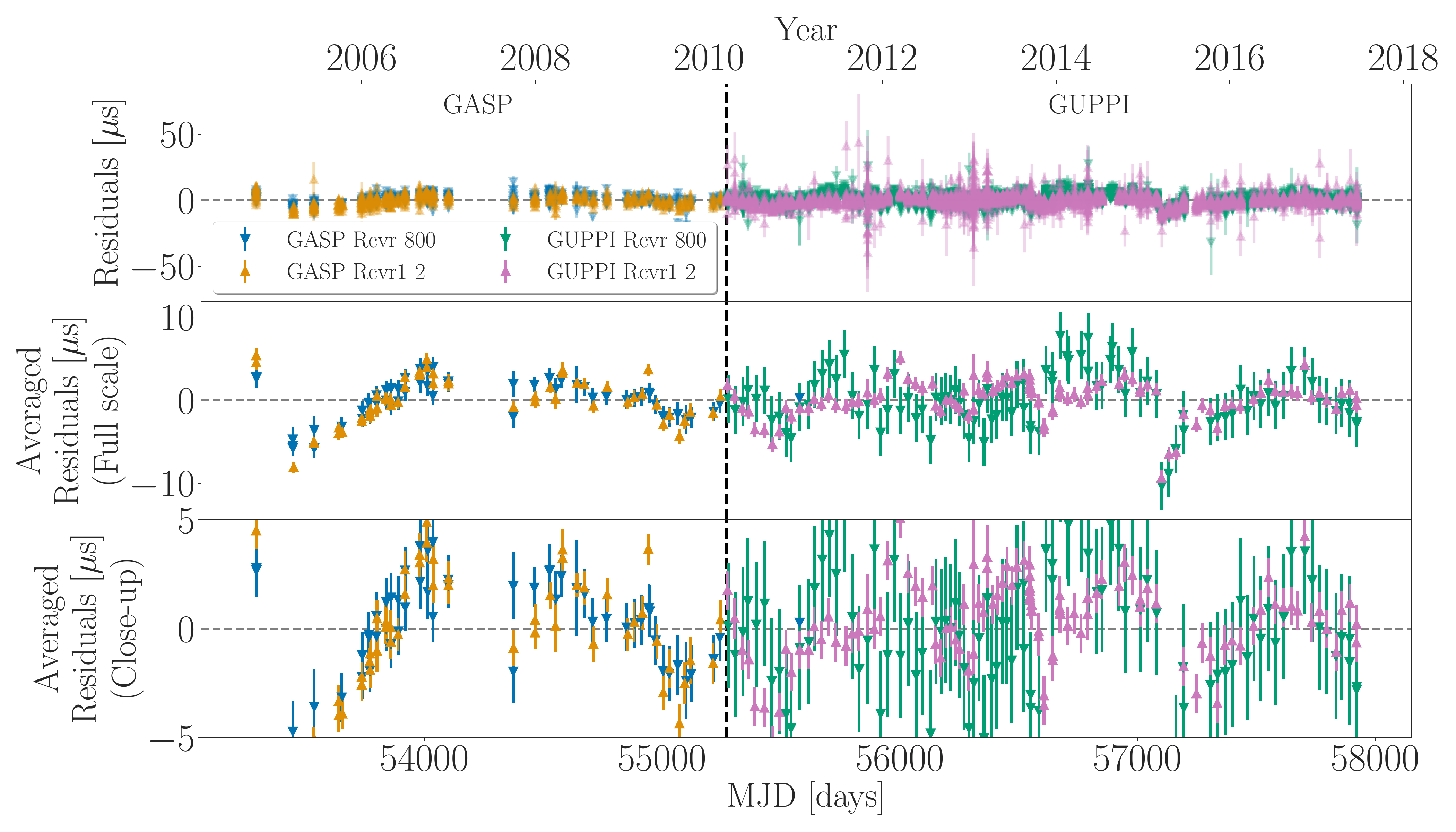}
\caption{Timing residuals for PSR J1643$-$1224 using NANOGrav's 12.5-yr data set \citep{Nanograv_12yr}. The predominant data acquisition backend instrument over any given time period is indicated at the top of each figure, and vertical dashed lines indicate the times at which instruments changed. Colored points indicate the receiver: Rcvr\_800 MHz (blue for GASP, orange for GUPPI) and Rcvr1\_2 (green for GASP, pink for GUPPI). Top panel: residual arrival times for all TOAs. Points are semi-transparent, and opaque regions arise from the overlap of many points. Middle panel: average residual arrival times shown in full scale. Each observation is composed of many simultaneously-obtained narrowband TOAs at different frequencies. Bottom panel: close-up of residuals around zero.
\label{fig:residuals}}
\end{figure*}

Two generations of pulsar backend processors were used for real-time coherent dedispersion and folding of the signal:

\begin{itemize}
    \item GASP was used from the start of the NANOGrav observing program  in 2004 until its decommissioning in 2012. It decomposed the signal into contiguous  4-MHz channels over a bandwidth of 64 MHz \citep{Ferdman_2008}.
    \item Starting in 2010 GASP was replaced by GUPPI, a wideband system that can process up to 800 MHz in bandwidth using smaller, 1.5625-MHz channels, and that significantly improved the timing precision relative to GASP \citep{10.1117/12.857666}. During the transition from GASP to GUPPI, precise measurements of time offsets between the instruments were made and included in the residual calculation \citep{Nanograv_12yr}.
\end{itemize}

The observations were calibrated and analyzed using standard pulsar processing techniques as implemented in the code \texttt{PSRCHIVE} \citep{2004PASA...21..302H} within the NANOGrav data reduction pipeline \citep{2018ascl.soft03004D}. In brief, the backend divides the telescope passband into narrow spectral channels, undertakes coherent dedispersion of the signals within each channel, and folds the resulting time series in real-time using a pulsar timing model. The data were thus transformed into folded pulse profiles as a function of time, pulsar phase, radio frequency, and polarization. These profiles have 2048 phase bins across the pulsar spin period, a frequency resolution of 4 MHz (GASP) or 1.5 MHz (GUPPI), and a time resolution (sub-integration time) of $1$ and $10\,\mathrm{s}$, respectively \citep{NANOGrav_11yr}. 

Care was taken to remove all artifacts that will result in a frequency-dependent systematic TOA bias. This includes removing image rejection artifacts that could arise from running two interleaved analog-to-digital conversion schemes if the gain of the two converters is not identical. Furthermore, the data set cleaning pipeline also involved systematically removing radio-frequency interference, excluding low-signal-to-noise ratio TOAs (see details in \citealt{NANOGrav_9yr}), removing outliers identified by Bayesian analysis of residuals (see details in \citealt{NANOGrav_11yr}), removing observations affected by calibration or digitization errors, and manual inspection of the data sets. The full details regarding data collection, calibration, pulse arrival-time determination, and noise modeling for the NANOGrav 12.5-year data set are provided in \cite{Nanograv_12yr}.

\begin{figure*}[ht!]
\epsscale{1.0}  
\plotone{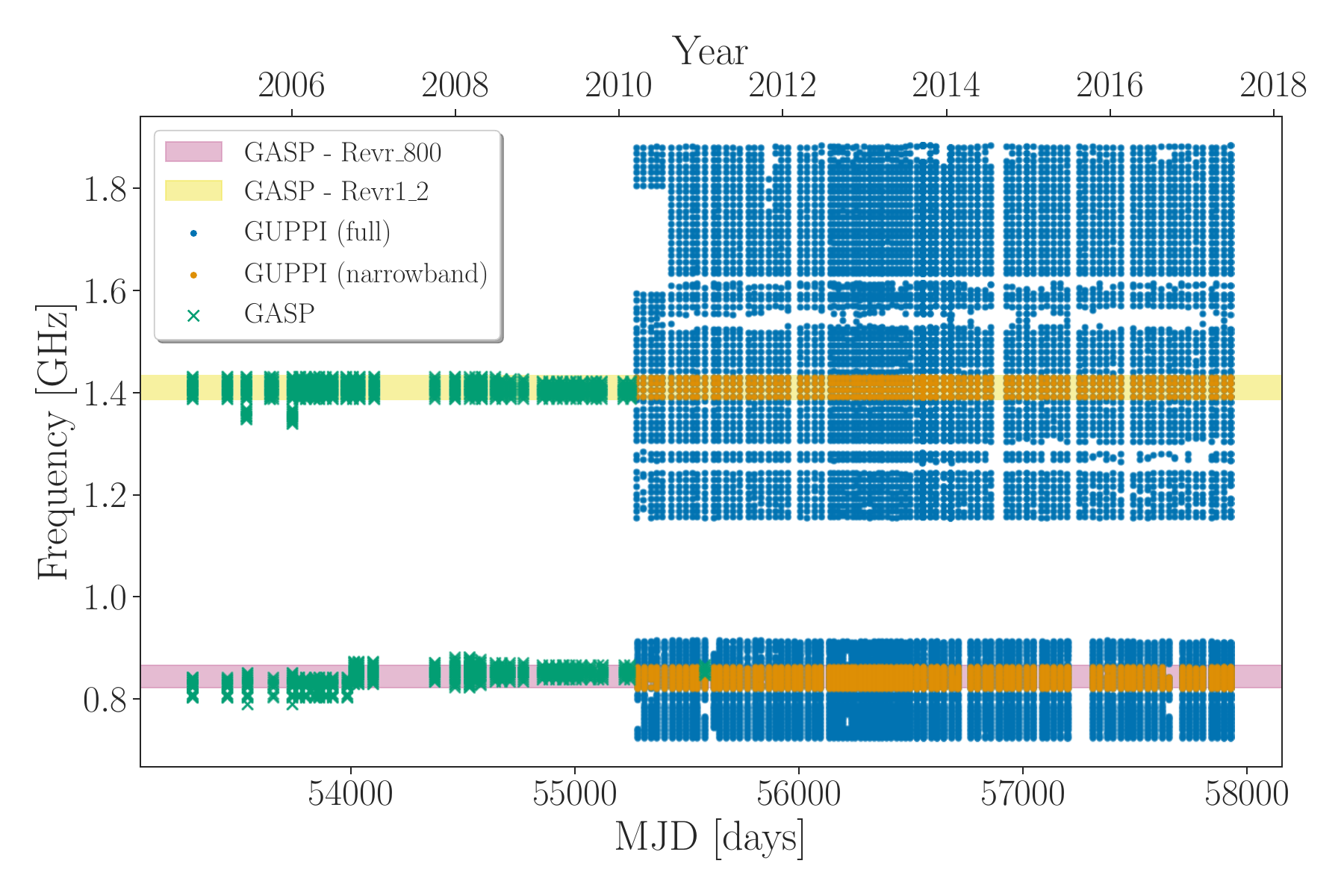}
\caption{Observing frequency as a function of the MJD for each observation in our data set. The typical frequency bands corresponding to each of GASP's receivers, Revr\_800 and Revr1\_2, are highlighted in purple and red. The full set of observations taken with GUPPI, covering a wide frequency range, is presented in purple; the subset of these observations that were used to emulate a narrowband data set is presented in red.
\label{fig:frequency_vs_MJD}}
\end{figure*}

\section{Narrowband and broadband fits \label{sec:results}}

\subsection{PSR J1643-1224}\label{subsec:j1643}

We expect that the effects of DM mis-estimations on timing residuals should be more clearly discernible in highly dispersed pulsars. Therefore, the main focus of our work are the observations of PSR J1643$-$1224, a 4.62~ms-period, high spin-down ($dP/dt = 1.85 \times 10^{-20}$) pulsar in a 147-day binary orbit with a white dwarf companion. This pulsar is of particular interest because lies behind the H\textsc{ii} region Sh 2$-$27, which has an inferred diameter of $0.034$~kpc assuming spherical symmetry \citep{Harvey_Smith_2011}. As a result, its pulses suffer high interstellar dispersion ($\mathrm{DM}=62.3~\mathrm{pc~cm^{-3}}$).  Furthermore, PSR J1643$-$1224 has been shown to have significant  scattering and profile shape variations \citep{Shannon_2016, 10.1093/mnras/stx580}. 

PSR~J1643$-$1224 has been observed by NANOGrav for over $12.7$~yr using the GBT with a nearly monthly cadence \citep{Nanograv_12yr}. During this time, its timing residuals have been reported to exhibit significant red noise (red noise spectral amplitude at $f=1~\mathrm{yr}^{-1}$, $A_{\mathrm{red}}=1.619 ~ \mathrm{\mu s} ~ \mathrm{yr}^{1/2}$) , which may include contributions from unmodeled interstellar-medium propagation effects  \citep{NANOGrav_11yr}. A dip in the timing residuals between 2015 February 21 and March 7 was found by \cite{2016ApJ...828L...1S}, which is associated with a sudden
change of pulse profile. When not modeled, this dip affects upper limits on
the stochastic gravitational-wave background (GWB), so it has been included in subsequent timing models.

In Fig.~\ref{fig:residuals} we present the timing residuals for PSR J1643$-$1224 using NANOGrav's 12.5-yr data release. We observe in the middle panel that during the time period when GASP was the predominant data acquisition backend, the epoch-averaged residuals are highly correlated and track each other. Most importantly, this trend is not present in the GUPPI data set. A possible explanation for this correlation are unaccounted offsets caused by DM mis-estimation in GASP's narrower bandwidth. In addition, we notice that the residuals in GUPPI's data set, especially those in the $800$ MHz band, generally exhibit larger uncertainties and a bigger spread than those taken with GASP in the same frequency band. These features are in agreement with the behavior predicted by Fig.~\ref{fig:cartoon} for a timing model that is affected by DM mis-estimations due to other chromatic effects: when we calculate residuals sampling a wider frequency range we expect some frequencies to be more heavily affected by such mis-estimations and, therefore, to exhibit larger residuals and uncertainties (orange curve in Fig.~\ref{fig:cartoon}), but those same frequencies might not be present when we sample narrow frequency ranges (green curve). These results demonstrate the need for an in-depth analysis of the effects of interstellar dispersion and the mis-estimations in its modeling on timing residuals.

\begin{deluxetable*}{lcccc}
\tabletypesize{\scriptsize}
\tablewidth{0pt} 
\tablecaption{\centering J1643$-$1224 data set}\label{tab:observations}
\tablehead{
\colhead{\multirow{3}{1.5cm}{\centering Backend}} & \colhead{\multirow{3}{1.5cm}{\centering Number of observations}} & \colhead{\multirow{3}{1.5cm}{\centering Data Span (MJD)}} & \colhead{\multirow{3}{1.5cm}{\centering Frequency Range (MHz)}} & \colhead{\multirow{2}{1.5cm}{\centering Used observations$^{(1)}$}} \\
} 
\startdata 
GASP & 1206 & 53291.91$-$55578.48 & 792$-$1432 & --- \\
GUPPI (full) & 11592 & 55275.26$-$57922.11 & 725.32$-$1882.81 & 9604 \\
GUPPI (narrowband) & 1813 & 55275.26$-$57922.11 & 822.33$-$1430.81 & 1511 \\
\enddata
\footnotesize{$^{(1)}$ After discarding observations in time windows that do not cover both frequency ranges.}\\
\end{deluxetable*}

For the purposes of this work, we started with a set of $11592$ observations taken with the GUPPI backend as early as 2010 until late 2017, covering a time baseline of $\sim 7.5$~yr. We have separated these observations into two data sets, the original and a modified version:
\begin{itemize}
    \item The full set of GUPPI broadband observations, which covers a bandwidth of $1157.49$~MHz. For practical purposes, we consider that the timing solution obtained from this data set provides the ``best estimation'' DM parameters that later on will be compared against those resulting from the narrowband approximation.
    \item In addition, we created an artificial set of narrowband observations by using \texttt{Astropy} \citep{astropy:2022} to filter out all the GUPPI broadband observations outside the frequency ranges of GASP's two receivers: Rcvr\_800 ($792$-$884$~MHz) and Rcvr1\_2 ($1340$-$1432$~MHz). In doing so, we emulate the data set that would have resulted from continuing to use GASP's bandwidth during the same time period.
\end{itemize}
These two sets of observations, alongside with the corresponding frequency ranges, are presented in Fig.~\ref{fig:frequency_vs_MJD}.

Using the time windows that are specified by the DMX parameters (see Sec.~\ref{sec:context}), we have grouped the observations into different time windows, each of them with observations up to 6 days apart. In order to accurately measure the pulsar’s dispersion properties on monthly timescales, and to account for any evolution in these frequency-dependent properties over time, we only considered windows that contain observations in both frequency bands and discarded all observations in windows that do not cover both bands. As a result, the set of broadband observations reduces to 9604 and the set of narrowband observations to 1511. Table \ref{tab:observations} summarizes the data sets used for this work.

\begin{figure*}[ht!]
\epsscale{1.0}
\plotone{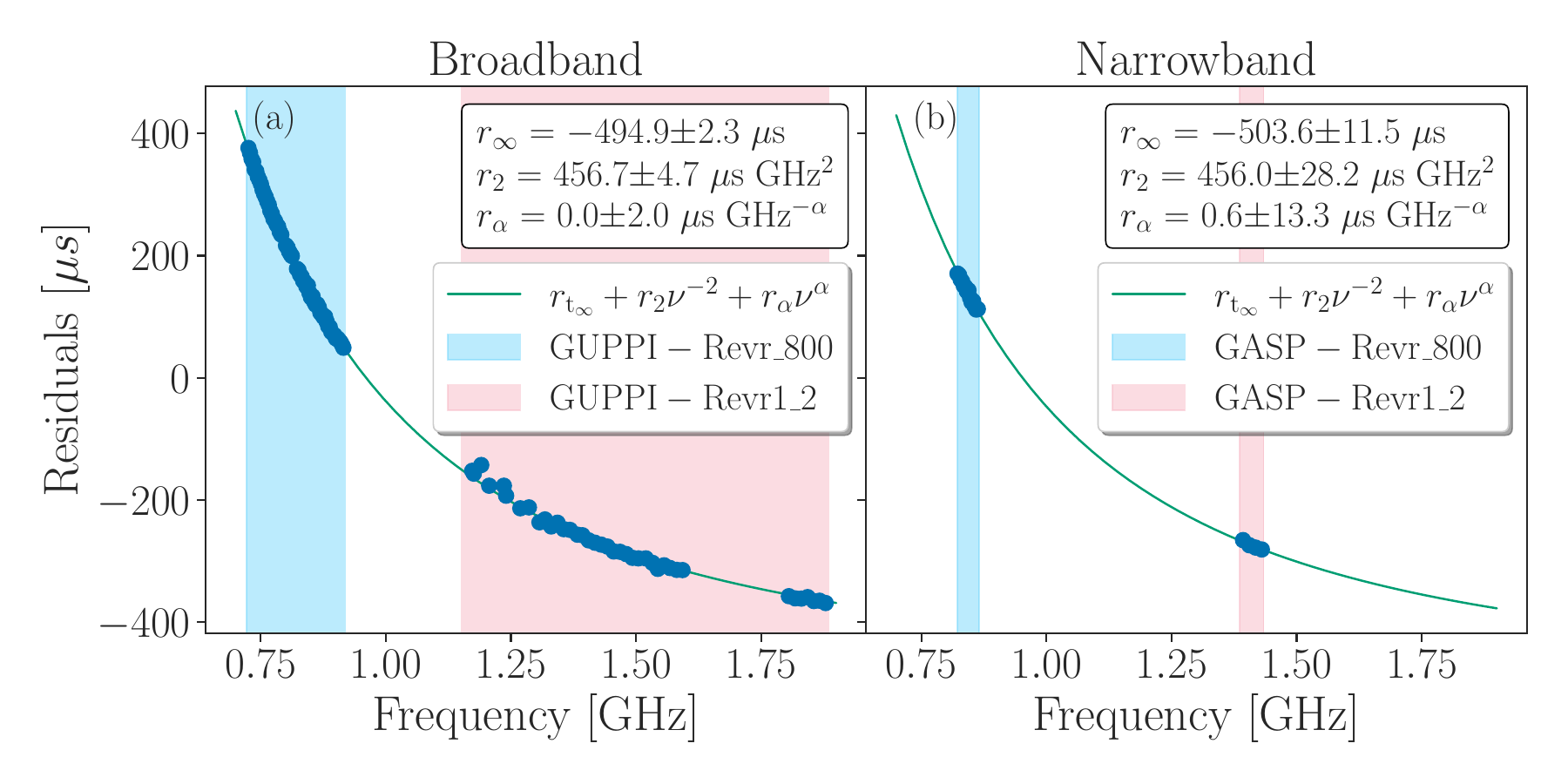}
\caption{Example of fitting the $r_{\infty}$, $r_{\mathrm{2}}$, and $r_{\mathrm{\alpha}}$ parameters using the residuals that are obtained subtracting the simplified timing model's predicted TOAs and either the broadband (panel (a)) or narrowband (panel (b)) set of observed TOAs of J1643$-$1224 that fall within the window from $\mathrm{MJD} = 55305.17896$ to $55307.17351$. The frequency ranges corresponding to the two GASP receivers, Revr\_800 and Revr1\_2, are shaded in green and red, respectively. Because the simplified timing model does not include DMX and additional chromatic corrections, the residuals are not distributed a round $R=0$ but follow a dispersion curve that can be fitted using Eq.~\ref{eq:disp_simplified}. The fitted values for this window and their errors are presented in the text boxes.
\label{fig:fits}}
\end{figure*}

\hspace{1cm}

\subsection{Biases in the timing parameters}

\begin{figure*}[htp]

\centering

\subfigure[J1643$-$1224]{\includegraphics[width=12.5cm]{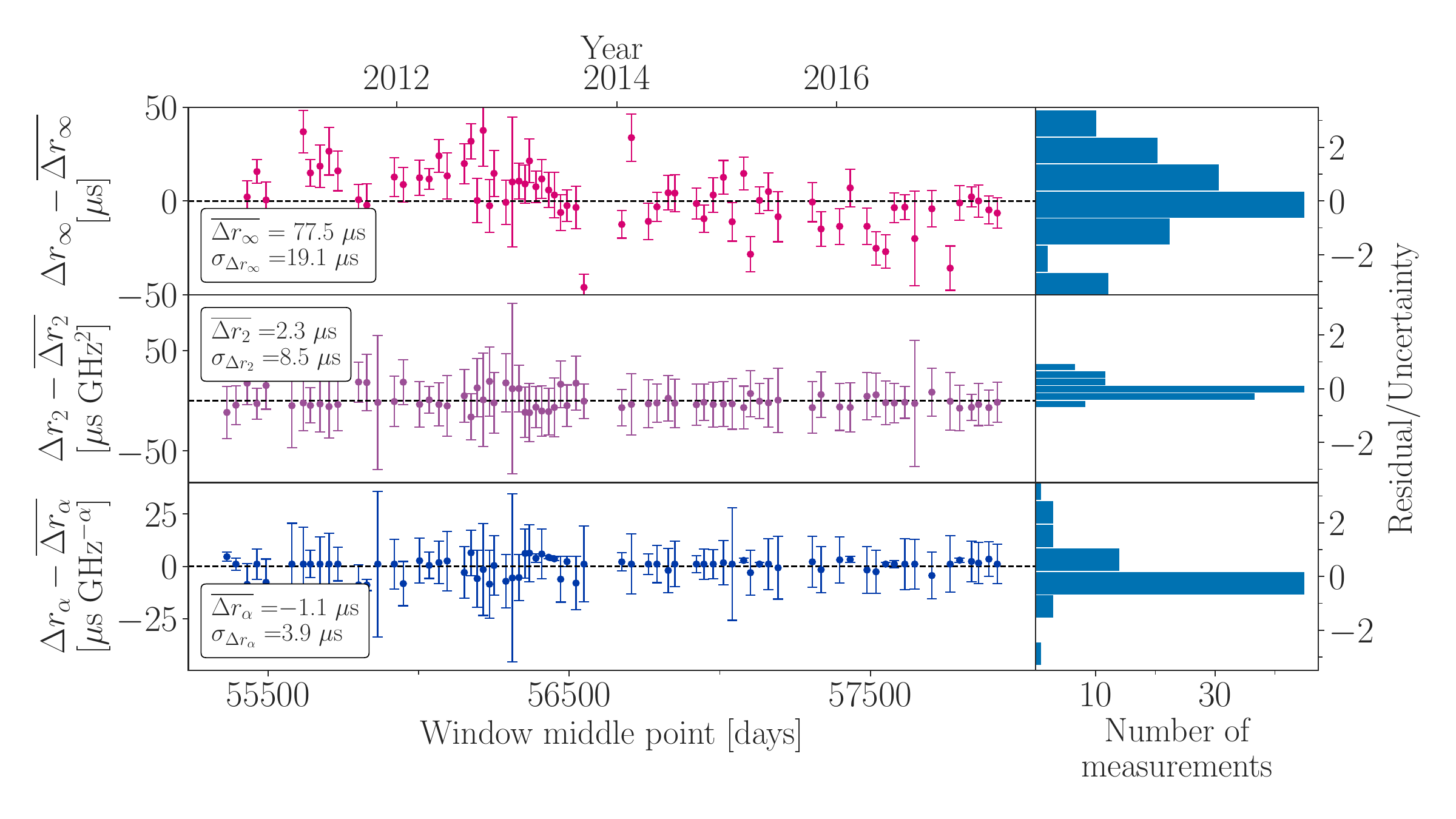}\label{fig:J1643_fits}}
\centering
\subfigure[J1744$-$1134]{\includegraphics[width=12.5cm]{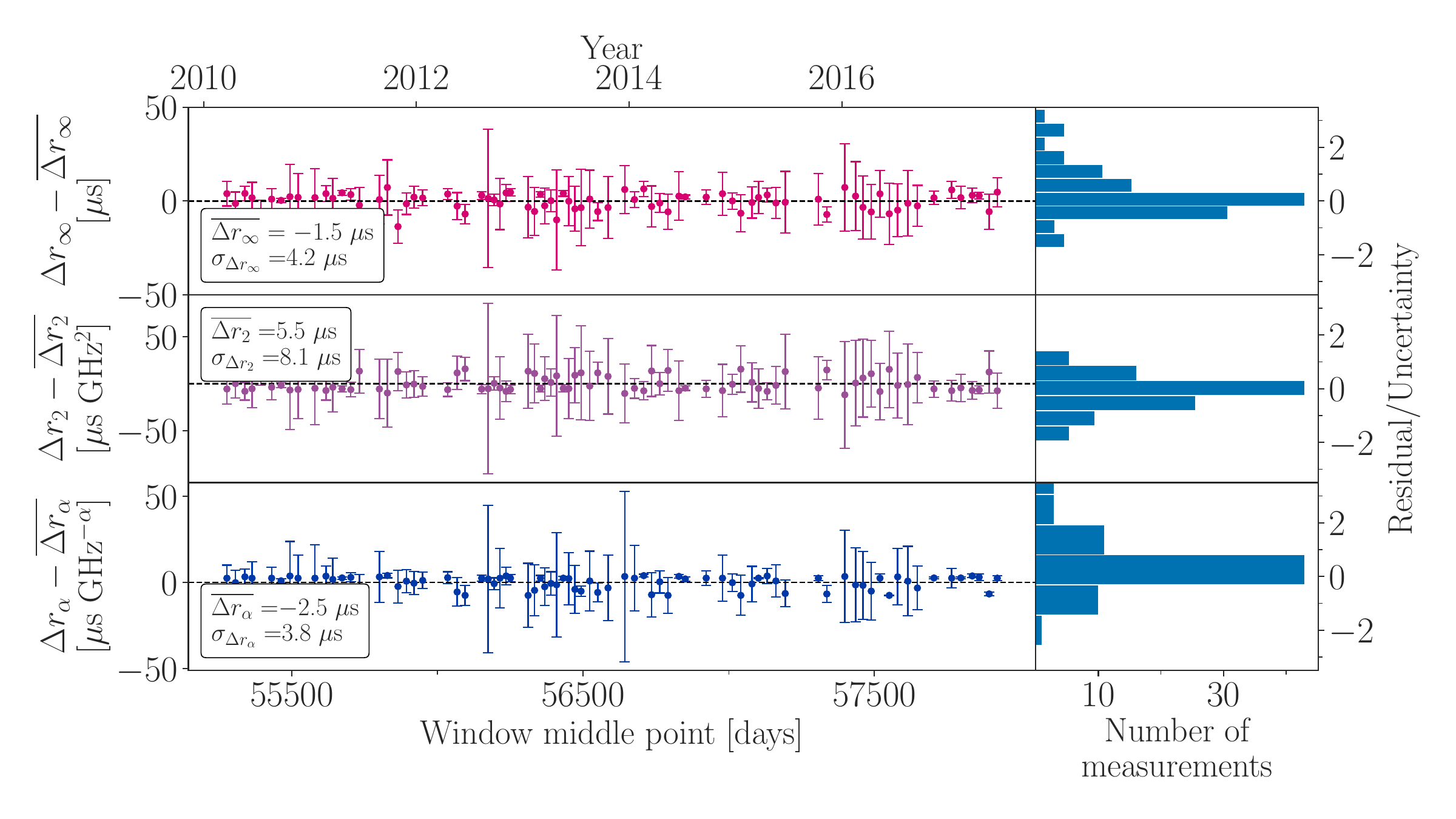}\label{fig:J1744_fits}}

\caption{Fitted parameters for J1643$-$1224 in figure (a) and J1744$-$1134 in figure (b). For each fitted parameter $r_{{k}}$ in Eq.~\ref{eq:disp}, each panel left represents the parameter residual $\Delta r_{{k, i}} = r_{{k, i}}^{\mathrm{BB}} - r_{{k, i}}^{\mathrm{NB}}$ between the values that were fitted at epoch $i$ using the broadband and the narrowband data sets. We also report the mean value of all the differences, $\overline{\Delta r_{k}}$, and the standard deviation of each set of residuals, $\sigma_{\Delta r_{k}}$. Since only differences are relevant, the mean value of the residuals is subtracted in each case. For each parameter, in the right panel we present histograms of the number of residuals divided by their corresponding fitting errors $ \varepsilon_{\Delta r_{k, i}}$ (see Eq.~\ref{eq:diff_error}).
\label{fig:parameter_fit}}\label{fig:differences}
\end{figure*}

To isolate the biases introduced in the timing parameters by fitting them using narrowband observations, we created a simplified timing model that includes corrections for long-term interstellar dispersion and frequency-dependent profile evolution (the $K \times \mathrm{DM} / \nu^2$ and $t_\mathrm{PE, \nu}$ terms in Eq.~\ref{eq:disp}, respectively) but ignores the short-scale variations in the DM that are normally corrected by the $\mathrm{DMX}$ parameters, as well as additional chromatic variations (the $K \times \mathrm{DMX} / \nu^2$ and $t_\mathrm{C, \nu}$ terms). This simplified timing model is loaded into \texttt{PINT} \citep{PINT}, a Python package used for pulsar timing and related activities. We then used this model to generate predicted TOAs at each of the observing dates and frequencies, which are then subtracted from the observed TOAs to produce timing residuals. As a result of ignoring the short-scale DM variations and additional chromatic delays, the residuals within each of the time will display exhibit a curve close to $\nu^{-2}$ as presented in Fig.~\ref{fig:fits}. We can then model the effects of the chromatic delays assuming the residuals can be described by
\begin{equation}\label{eq:disp_simplified}
    r_{i} ( \nu ) = r_{\infty, i} + r_{\mathrm{2}, i} \nu^{-2} + r_{\mathrm{\alpha}, i} \nu^{-\alpha},
\end{equation}
where $r_{\mathrm{\infty}, i}$ is the residual expected at epoch $i$ if no chromatic effects were present, $r_{\mathrm{2}, i} = K \times \mathrm{DMX}_{i}$ is the correction for short-scale DM variations, $r_{\mathrm{\alpha}, i}$ is the correction due to a scattering-based delayed, corresponding to $\alpha = 4.4$. We used \texttt{LMFIT} \citep{LMFIT} to fit Eq.~\ref{eq:disp_simplified} to the residuals within each time set, thereby obtaining best-fit values and uncertainties for $r_{\mathrm{\infty}}$, $r_{\mathrm{2}}$, and $r_{\mathrm{\alpha}}$. Such a fit is data set-dependent and any unaccounted biases in the observations, such as DM mis-estimations, will result in biased fitted parameters.

\begin{figure*}[btp]
\epsscale{0.9}
\plotone{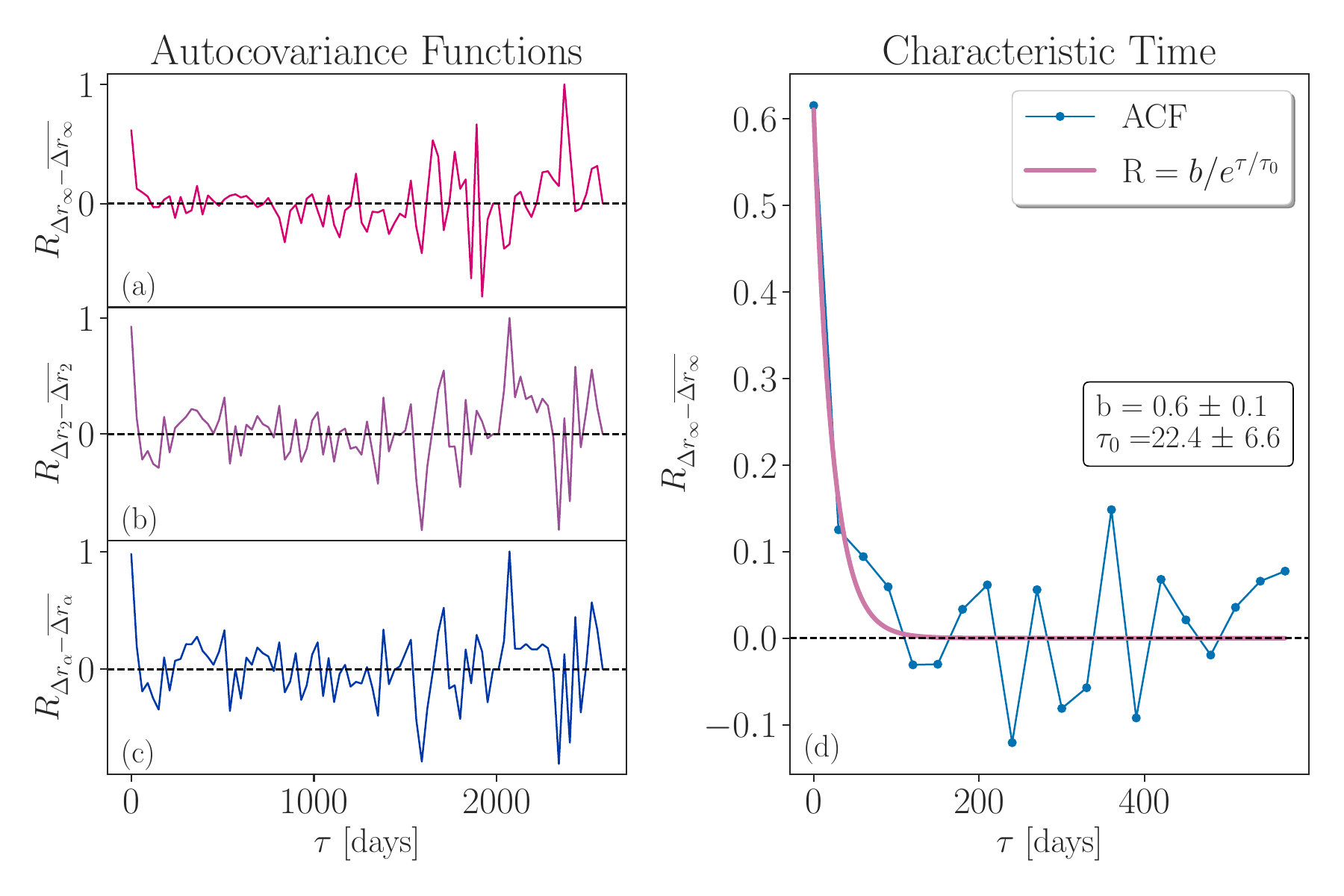}
\caption{Autocovariance functions for PSR J1643$-$1224. Panels (a), (b), and (c) show the autocovariances for the values of $\Delta r_{\infty} - \overline{\Delta r_{\infty}}$, $\Delta r_\mathrm{2} - \overline{\Delta r_\mathrm{2}}$, and $\Delta r_\mathrm{\alpha} - \overline{\Delta r_\mathrm{\alpha}}$ that are  presented in Fig.~\ref{fig:J1744_fits}. In panel (d) we fit an exponential $R = b/e^{\tau / \tau_{0}}$ function to the first 20 datapoints in the autocovariance function for $\Delta r_{\infty} - \overline{\Delta r_{\infty}}$, resulting in a characteristic time of $\tau_{0} = 22.4 \pm 6.6$~days within which the values of $\Delta r_{\infty}$ are correlated.}
\label{fig:autocovariance}
\end{figure*}

This process was repeated using the narrowband and the broadband data sets separately (see Sec.~\ref{sec:observations}). As a result, for each time window at epoch $i$, we obtained two sets of $\{r_{\infty, i}, r_{\mathrm{2}, i}, r_{\mathrm{\alpha}, i} \}=\{r_{k, i}\}_{k=\infty, 2, \alpha}$ values with their corresponding fitting errors $\varepsilon_{r_{k, i}}$: one set resulting from using the broadband (BB) data set and another from using the narrowband (NB) one. For each $r_{k,i}$ we computed the parameter residuals 

\begin{equation}\label{eq:diff}
    \Delta r_{k, i} = r_{k, i}^{\rm BB} - r_{k, i}^{\rm NB} \quad , \quad k=\infty, 2, \alpha
\end{equation}

\noindent between the values fitted using the broadband and the narrowband observations in each epoch. The error associated with each $\Delta r_{k, i}$ is given by

\begin{equation}\label{eq:diff_error}
    \varepsilon_{\Delta r_{k, i}} = \sqrt{(\varepsilon^{\mathrm{BB}}_{r_{k, i}})^2+(\varepsilon^{\mathrm{NB}}_{r_{k, i}})^2}
\end{equation}

\noindent where $\varepsilon^{\mathrm{BB}}_{r_{k, i}}$ is the fitting error from fitting $r_{k, i}$ using the BB data and $\varepsilon^{\mathrm{NB}}_{r_{k,i}}$ is the fitting error from using the NB data.
Moreover, since we are only interested in the relative differences between both sets of observations, we have subtracted the mean value of all the differences, $\overline{\Delta r_{k}}$, from each of the differences $\Delta r_{k, i}$. 

In the left panels of Fig.~\ref{fig:J1643_fits} we plot the resulting values of $\Delta r_{k, i} - \overline{\Delta r_{k}}$ with their errors as a function of the MJD at the middle of the window. If the frequency bandwidth played no role in modeling interstellar dispersion, we would expect to obtain $\Delta r_{k, i} = 0$ for all the parameters $k$ and all epochs $i$. However, the fact that we find non-zero deviations between the broadband and narrowband sets of fitted parameters is indicative that using narrower frequency bandwidth leads to mis-estimations in modeling these parameters for PSR~J1643$-$1224. In order to quantify such mis-estimations, we calculate the standard deviation for each set of parameter differences, which we will hereby use as a measurement of the error introduced in the residuals due to incomplete modeling of interstellar dispersion. The largest offset, corresponding to $r_{\infty}$ is $\sigma_{\Delta r_{\infty}}=22.2~\mu$s.

For each fitted parameter, in the right panels of Fig. \ref{fig:J1643_fits} we plot histograms of the residuals $\Delta r_{k, i} - \overline{\Delta r_{k}}$ divided by their corresponding fitting errors $ \varepsilon_{\Delta r_{k, i}}$. We observe that for $r_{2}$ and $r_{\alpha}$ the residuals are consistent with zero, which is indicative of a small offset introduced in the fit of these parameters when using narrowband data. However, the histogram for $r_{\infty}$ reveals a skewed distribution (significantly more noticeable for J$1643-1224$) which suggests a systematic offset in the estimations of the infinite frequency arrival times.


\subsection{Autocovariance Functions}

Next, in order to study variations in the behaviour of the ISM, we analyze whether the chromatic delay exhibits correlations in time. In that case, we expect such a pattern to be revealed in the autocovariance function (ACF) of this quantity. Therefore, we compute the ACF of  $\Delta r_{k, i} - \overline{\Delta r_{k}}$ (see Eq.\ref{eq:diff}) for each parameter $r_{k}$ between consecutive time windows.
This process involves correlating a signal $f(t)$ with a delayed copy $f(t+\tau)$ of itself as a function of delay the $\tau$. However, since we have a discrete and irregularly sampled signal, we calculated a binned ACF. For a given delay $\tau$, we averaged the correlation $f(t_m) f(t_n)$ between all point pairs that are spaced apart by a time difference $t_m - t_n$ within the range $\tau \pm \Delta \tau /2$, where $\Delta \tau = 30$~days is the bin width and the normalization constant is given by the number of point pairs that satisfy this condition, $N_{\tau}$. More precisely:

\begin{equation}
    R(\tau) = \frac{1}{N_{\tau}} \underbrace{\sum_{t_m} \sum_{t_n}}_{|t_m - t_n| \in [\tau \pm \Delta \tau / 2]} f(t_m) f(t_n)
\end{equation}

\noindent The resulting ACFs for each of the fitted parameters are shown in Fig.~\ref{fig:autocovariance} panels (a), (b), and (c).

As a byproduct of the ACFs, we also found the characteristic time $\tau_{0}$ over which the values of $\Delta r_{k, i} - \overline{\Delta r_{k}}$ are correlated. For this purpose, we assume that the ACF follows an exponential function given by $R(\tau)= b e^{-\tau/\tau_{0}}$, where $b$ and $\tau_{0}$ are free parameters that we fit to the ACF. In doing so, we find a characteristic time of $\tau_{0}=22.4 \pm 6.6$~days, which corresponds to roughly the observing cadence. The fitted function and the derived parameters are presented in Fig.~\ref{fig:autocovariance} panel (d).


\begin{figure*}[ht!]
\centering
\includegraphics[width=0.75\paperwidth]{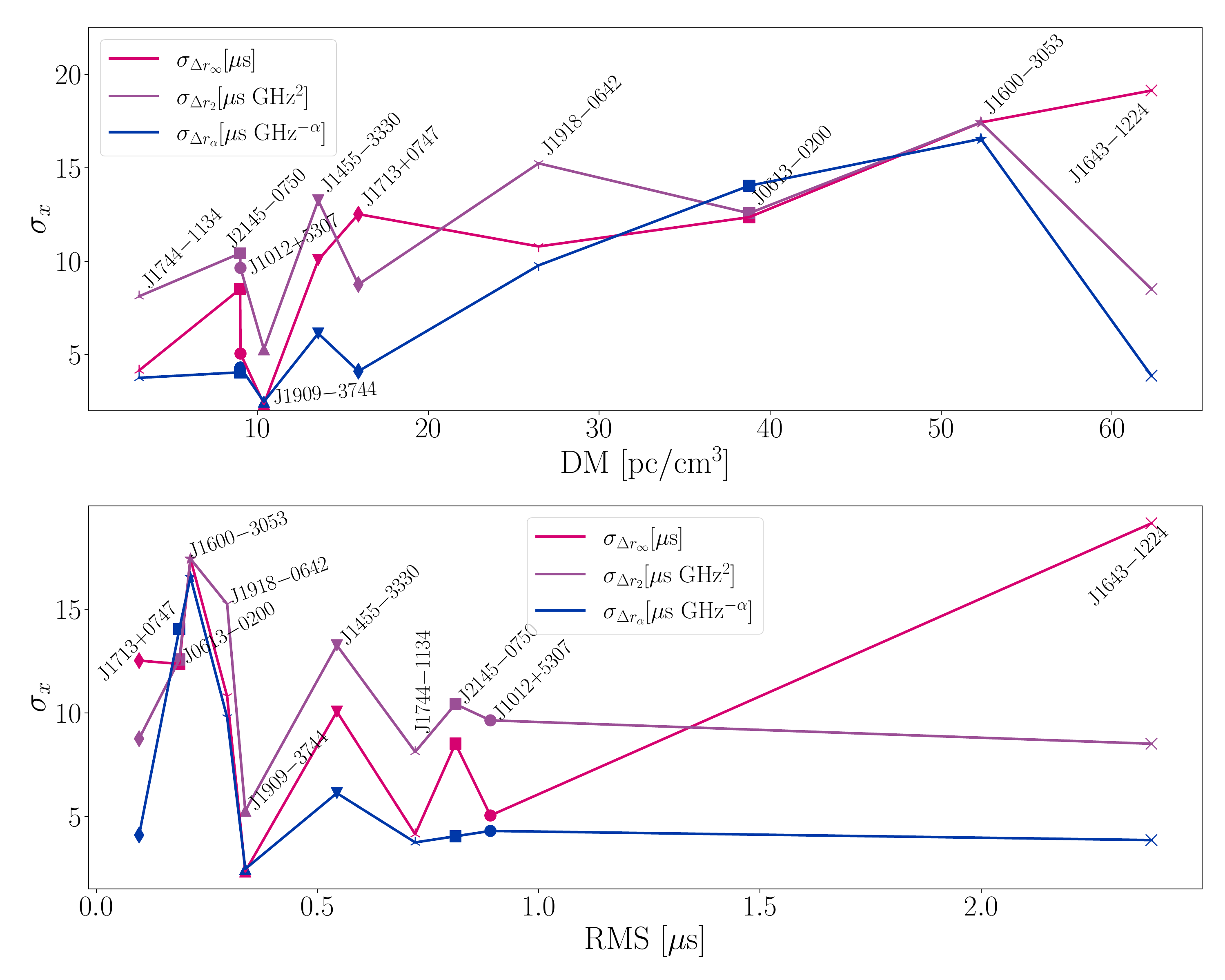}
\caption{Top panel: Standard deviation of $\Delta r_{\infty}$, $\Delta r_{2}$, and $\Delta r_{\alpha}$ for the different pulsars we have included in this study, as a function of the pulsar DM. Each pulsar is represented with a different marker; each parameter is represented with a different color. The DM values were obtained from \cite{Limits_on_the_Stochastic_GW_Background}. Bottom panel: Standard deviation of $\Delta r_{\infty}$, $\Delta r_{2}$, and $\Delta r_{\alpha}$ as a function of the weighted root-mean-square (RMS) of post-fit timing residuals. The RMS values were obtained from \cite{2021ApJS..252....5A}.}
\label{fig:all_pulsars}
\end{figure*}

\subsection{Other Pulsars}

For completeness, we repeated this analysis for a sample of other pulsars observed by NANOGrav using GBT. This includes high-DM pulsars such as PSR J1600$-$3053 ($52.33~\mathrm{pc}~\mathrm{cm}^{-3}$, \citealt{2007ApJ...656..408J}) and PSR J1744$-$1134, which is one of the lowest-DM pulsars observed by NANOGrav ($3.14~\mathrm{pc}~\mathrm{cm}^{-3}$, \citealt{Limits_on_the_Stochastic_GW_Background}). 

The broadband-narrowband offsets in $r_{\infty}$, $r_\mathrm{2}$, and $r_\mathrm{\alpha}$ for PSR J1744$-$1134 are presented in Fig.~\ref{fig:J1744_fits}.  We find that this pulsar yields smaller mis-estimations  than those obtained for PSR J1643$-$1224. In particular, the mis-estimation in the infinite-frequency time of arrival, $\sigma_{\Delta r_{\infty}}$, is $4.4~\mu$s, which is  $
\sim 4.5$ times smaller than the value obtained for PSR J1643$-$1224. On the other hand, for J1600$-$3053 we find $\sigma_{\Delta r_{\infty}}=14.12~\mu$s, which is only $~\sim 1.3$ times smaller than $\sigma_{\Delta r_{\infty}}$ for PSR J1643$-$1224..

In order to study whether these differences are a result of J1744$-$1134 being less affected by interstellar dispersion, we also extend this analysis to other pulsars covering a broad range of DM values. The surveyed pulsars and the resulting values are summarized in Fig.~\ref{fig:all_pulsars}. For each pulsar, we present the mis-estimation $\sigma_{\Delta r_{k}}$ as a function of the pulsar $\mathrm{DM}$ (top panel) and as a function of the RMS of its timing residuals (bottom panel), given that the former is another observational property of the pulsar that could have an effect on the mis-estimations. The results show no obvious dependence of the mis-estimation $\sigma_{\Delta r_{k}}$ with the pulsar $\mathrm{DM}$ or its residual RMS. If we set an arbitrary cut-off value close to the middle of our DM range, at $\mathrm{DM}=30$~pc~cm$^{-3}$, we find that the average mis-estimation for lower-DM pulsars is $6.92~\mu$s, and for higher-DM pulsars it is $12.01~\mu$s. Broadly speaking, it can then be expected that high-DM pulsars will be generally more affected by dispersion mis-estimations in narrowband observations. However, the exact behavior is dependent on the specifics of the ISM in the pulsar line-of-sight, and no precise dependence of $\sigma_{\Delta r_{k}}$ as a function of the pulsar $\mathrm{DM}$ or its timing RMS can be established at this point.


\section{Conclusions \label{sec:conclusions}}

In this work we have quantified the systematic biases in the timing parameters describing chromatic delays that result from sampling the pulsar frequency spectrum using narrowband radio receivers. The effect is dependent on the DM of a given pulsar, but for a high-DM pulsar such as PSR J1643$-$1224 we find an offset as large as $22.2~\mu$s. Since timing models depend on these parameters to calculate the timing residuals, this analysis provides an estimate of the error that is consequently introduced in the residuals as a result of DM mis-estimations. Moreover, for J1643$-$1224 we find that the errors in observations at different epochs are correlated within $\sim1$ month of each other. For low-DM pulsars such as J1744$-$1134, the bias in the timing parameters reduces to $8.1~\mu$s. However, in Fig.~\ref{fig:all_pulsars} we do not find a clear linear dependence of the error as a function of the DM; instead, the exact offset is highly dependent on the pulsar properties. Nevertheless, for a typical DM value ($\leq 40$~pc~cm$^{-3}$) the systematic offset will be $\sim5~\mu$s.

The most immediate application of these results will be quantifying previously unaccounted error in NANOGrav's legacy observations, in order to incorporate them into the current data set. Legacy data comprise years of observations that are already available, and which could significantly strengthen current evidence for a detection of the long-period GW background \citep{NG15GWB}. For PTAs in the weak-signal regime (where the lowest frequency of the stochastic background power spectrum is below the white noise level), the signal-to-noise ratio scales with time as roughly the $4^\mathrm{th}$ power of the time baseline \citep{Siemens_2013}. Therefore, incorporating even a few years of legacy observations will result in an exponential increase in sensitivity. For PTAs in the strong-signal regime, the signal-to-noise ratio will scale as the square root of the time baseline.

The new generation of broadband radio receivers will substantially improve our estimations of the dispersion parameters. In particular, we find that DM mis-estimations are significantly mitigated by using broadband radio receivers, such as VEGAS at GBT. Moreover, the upcoming Deep Synoptic Array-2000 (DSA-2000) telescope, which will produce pulsar observations across the entire $0.7$–$2$~GHz band \citep{2019BAAS...51g.255H}, is expected to provide major improvements in current ISM models and to significantly reduce the residual errors due to biased DM parameters. Repeating the analysis presented in this work using DSA-2000 broadband observations could potentially contribute to better constraining DM mis-estimations.

Even as we move towards ultrawideband systems, this work has the potential to improve already existing narrowband data sets. In particular, we expect DM-induced biases to still be prevalent in GASP- and GUPPI-based observations (see Fig.~\ref{fig:residuals}). Narrowband observations are also predominant in, for example, the European Pulsar Timing Array data release  1.0 \citep{10.1093/mnras/stw483}, which subsequently played a major role in the International Pulsar Timing Array (IPTA) data release 2 \citep{2022MNRAS.510.4873A}. In particular, the Effelsberg Radio Telescope processes observations with a bandwidth up to $112$~MHz \citep{Backer_1997}, the Lovell Radio Telescope up to 128 MHz, the Nançay Radio Telescope uses a coherent de-dispersion backend of the same family as GBT's ASP-GASP with a bandwidth of either $64$ or $128$ MHz, and the Westerbork Synthesis Radio Telescope uses either $10$, $80$, or $160$~MHz of bandwidth. Therefore, quantifying the DM mis-estimations introduced by these narrowband systems is of utmost relevance for the search of an isotropic stochastic GW background in the IPTA data set.

These results are also of interest for other radio-astronomical facilities that are currently utilizing narrowband receivers. For example, the Argentine Institute of Radio Astronomy has two single-dish telescopes capable of performing daily pulsar monitoring, which could contribute to improving the IPTA's sensitivity to single sources of GWs \citep{Lam_2021}. However, their observations use instantaneous bandwidths of $112$~MHz and $56$~MHz \citep{2020A&A...633A..84G, 2023MNRAS.521.4504Z}. Therefore, using this type of analysis to account for DM mis-estimations might prove of foremost importance in achieving the timing precision required for contributing to future IPTA data sets.

\vspace{5mm}
\facilities{Green Bank Observatory (GBO).}


\software{PINT \citep{PINT}, PyPulse \citep{2017ascl.soft06011L}, LMFIT \citep{LMFIT}, Astropy \citep{astropy:2022}.}

\begin{acknowledgments}
\textit{Acknowledgments and author contributions}: S.V.S.F. undertook the analysis, developed the code pipeline, and prepared the ﬁgures, tables, and the majority of the text. M.T.L. developed the mathematical framework for this work, selected the analyzed data set, assisted with the preparation of the manuscript, provided advice interpreting the results, and supervised the project development. MTL, ZA, HB, PRB, HTC, MED, PBD, TD, JAE, RDF, ECF, EF, NGD, PAG, MJ, DRL, RSL, MAM, CN, DJN, TTP, SMR, RS, IHS, KS, JKS, SJV, and WWZ all ran observations and developed timing models for the NANOGrav 12.5-year data set.\textbf{}

We thank Olivia Young, James Cordes, Joris Verbiest, Joseph Lazio, and NANOGrav's Noise Budget and Timing Working Groups for their valuable input and feedback on the manuscript. 

The NANOGrav Collaboration receives support from National Science Foundation (NSF) Physics Frontiers Center award Nos. 1430284 and 2020265. The Green Bank Observatory and National Radio Astronomy Observatory  are facilities of the NSF operated under cooperative agreement by Associated Universities, Inc. M.T.L., T.D., and S.V.S.F. are supported by an NSF Astronomy and Astrophysics Grant (AAG) award number 2009468. S.V.S.F. acknowledges partial support by the NASA New York Space Grant, and by the Out to Innovate Career Development Fellowship for Trans and Non-binary People in STEM 2023. 
P.R.B. is supported by the Science and Technology Facilities Council, grant number ST/W000946/1.
Support for H.T.C. is provided by NASA through the NASA Hubble Fellowship Program grant \#HST-HF2-51453.001 awarded by the Space Telescope Science Institute, which is operated by the Association of Universities for Research in Astronomy, Inc., for NASA, under contract NAS5-26555.
M.E.D. acknowledges support from the Naval Research Laboratory by NASA under contract S-15633Y.

E.C.F. is supported by NASA under award number 80GSFC21M0002.
The Flatiron Institute is supported by the Simons Foundation.
D.R.L. and M.A.M. are supported by NSF \#1458952.
M.A.M. is supported by NSF \#2009425.
The Dunlap Institute is funded by an endowment established by the David Dunlap family and the University of Toronto.
T.T.P. acknowledges support from the Extragalactic Astrophysics Research Group at E\"{o}tv\"{o}s Lor\'{a}nd University, funded by the E\"{o}tv\"{o}s Lor\'{a}nd Research Network (ELKH), which was used during the development of this research.
N.S.P. was supported by the Vanderbilt Initiative in Data Intensive Astrophysics (VIDA) Fellowship.
S.M.R. and I.H.S. are CIFAR Fellows.
Pulsar research at UBC is supported by an NSERC Discovery Grant and by CIFAR.
S.J.V. is supported by NSF award PHY-2011772.
\end{acknowledgments}






\bibliographystyle{mnras}
\bibliography{main}



\end{document}